\documentclass[preprint,11pt]{article}

\usepackage{bm}
\usepackage{epsfig}
\usepackage{color}
\usepackage{amssymb}
\usepackage{amsmath}
\usepackage{cite}

\begin{document}
\title{Dynamics and Statistics of the Fermi--Pasta--Ulam $\beta $--model with different ranges of particle interactions}

\author{
\textbf{Helen Christodoulidi$^{1}$, Tassos Bountis$^{1,2}$, Constantino Tsallis$^{3,4}$ and Lambros Drossos$^{5}$}\\
$^{1}$Center for Research and Applications of Nonlinear 
Systems,\\ University of Patras, GR-26500 Patras, Greece. \\
$^2$Department of Mathematics, Nazarbayev University,\\ Kabanbay-Batyr  53, 010000 Astana, Republic of Kazakhstan\\
$^3$Centro Brasileiro de Pesquisas Fisicas and \\
National Institute of Science and Technology for Complex Systems,\\ 
Rua Xavier Sigaud 150, 22290-180 Rio de Janeiro-RJ, Brazil\\
$^4$Santa Fe Institute, 1399 Hyde Park Road, Santa Fe, NM 87501, USA\\
$^5$High Performance Computing Systems and Distance Learning Lab,\\
Technological Educational Institute of Western Greece,\\ GR-26334 Patras, Greece}

\date{\today}%

\maketitle

\begin{abstract}
In the present work we study the Fermi--Pasta--Ulam (FPU) $\beta $--model involving long--range interactions (LRI) 
in both the quadratic and quartic potentials, by introducing two independent exponents $\alpha_1$ and $\alpha_2$ respectively, which make the {forces decay} with distance $r$. Our results demonstrate that weak chaos, in the sense of decreasing Lyapunov exponents, and $q$--Gaussian  probability density functions (pdfs) of sums of the momenta, occurs only when long--range interactions are included in the quartic part. More importantly, for $0\leq \alpha_2<1$, we obtain extrapolated values for $q \equiv q_\infty >1$, as $N\rightarrow \infty$, suggesting that these pdfs persist in that limit. On the other hand, when long--range interactions are imposed only on the quadratic part, strong chaos and purely Gaussian pdfs are always obtained for the momenta.  We have also focused on similar pdfs for the particle energies and have obtained $q_E$-exponentials (with $q_E>1$) when the quartic-term interactions are long--ranged,  otherwise we get the standard Boltzmann-Gibbs weight, with $q=1$. The values of $q_E$ coincide, within small discrepancies, with the values of $q$ obtained by the momentum distributions.

\end{abstract}



\section{Introduction}
\label{intro}

 
In recent years, many authors have examined the effect of long--range interactions on the dynamics of multi--dimensional Hamiltonian systems
\cite{Ruffo95,Ruffo98,AnteneodoTsallis1998,Dauxois2002,Tarasov06,Ginelli1,Ginelli2,Thanos}. Perhaps the best known example in this class is the so--called Hamiltonian Mean Field model, where the maximal Lyapunov exponent (MLE) was shown numerically to decrease with increasing number of degrees of freedom $N$, according to a specific power law \cite{Ruffo98,AnteneodoTsallis1998,Ginelli1,Ginelli2}. More recently, another famous example in this category, the FPU $\beta $--Hamiltonian was studied in the presence of long--range interactions. In the complete absence of harmonic terms the MLE appears to vanish in the thermodynamic limit \cite{Bagchi}. Moreover, when harmonic terms are included in the potential, a similar behavior of the MLE is observed \cite{CTB}, which nevertheless tends to saturate to a non--zero value above a characteristic size $N$. 

It is the purpose of the present paper to investigate more thoroughly the FPU $\beta $--model from this point of view, by studying the effect of the interactions through two parameters $\alpha_1$ and $\alpha_2$ introduced in the quadratic and quartic terms of the potential respectively. In so doing, we are able to identify domains of strong and weak chaos, by examining whether probability density functions (pdfs) of sums of the momenta obey Boltzmann Gibbs (BG) statistics or not.

The maximal Lyapunov exponent and other indicators of local dynamics \cite{GALI} provide useful tools for chaos detection, but are not well suited for distinguishing between different degrees of weak vs. strong chaos \cite{bountis_book}. For example, if a given orbit is chaotic, its MLE is expected to converge to a positive value. However, if the orbit is trapped for a long time near islands of regular motion, the MLE does not quickly converge and when it does, one cannot tell from its value whether the dynamics can be described as weakly or strongly chaotic.  

Now, long--range systems are known to possess long--living quasi--stationary states (QSS) 
\cite{Leo1,Leo2,Leo3,PluchinoRapisardaTsallis2007,PluchinoRapisardaTsallis2007b,CirtoAssisTsallis2013,uma1,uma2,uma3,uma4,uma5,
uma6,uma7,TsallisPlastinoAlvarezEstrada2009}, whose statistical properties are very different from what is expected within the framework of classical BG thermostatistics \cite{Gibbs1902}. More specifically, when one studies such QSS in the spirit of the central limit theorem, one finds that the pdfs of sums of their variables are well approximated  by $q$--Gaussian functions (with $1<q<3$) or $q$--statistics \cite{ABB2011,Tsallis2009,Tsallis1988, GellMannTsallis2004,Tsallis2014}. These pdfs last for very long times beyond which they are expected to tend to the $q=1$ case of pure Gaussians and BG thermal equilibrium. Thus, we will treat the index $q$ as a measure of the ``distance'' from a Gaussian, and study its time evolution to identify when a ``phase transition'' will occur from weak chaos and $q>1$--statistics to strong chaos and BG thermostatistics.   

In this context, it becomes highly relevant to examine the effect of the range of the interactions on the lifetime of a QSS, and hence the duration of weakly chaotic dynamics. To this end, we recently introduced and studied numerically a generalization of the  FPU $\beta$--model, in which we varied the interaction range by multiplying the quartic terms of the potential by coupling constants that decay with distance as $r^{-\alpha}$ \cite{CTB}. The pdfs of the time--averaged momenta were thus found to be well approximated by $q$--Gaussians with $q>1$, when the range is long enough (i.e. $\alpha<1$). This, however, lasts up to a crossover time $t=t_c$ at which $q$ starts to decrease monotonically to 1, reflecting the transition from $q$--statistics to BG thermostatistics. 

In the present paper, we extend our study and investigate additional properties connected with the occurrence of weakly chaotic QSS, taking a closer look at their dynamics as well as associated statistics. In particular, we consider the FPU $\beta$--chain \cite{FPU} of $N$ particles, whose potential includes harmonic as well as quartic interactions, and employ two different exponents $\alpha_1$ and $\alpha_2$, for the $r^{-\alpha}$ coupling constants of the quadratic and quartic terms respectively. Furthermore, we vary these exponents independently to investigate their effect on the thermostatistics of the orbits at increasingly long times. A recent study on long--range interactions applied only on the harmonic terms can be found in \cite{Ruffo15}.

In Section \ref{sec2}, we write the Hamiltonian of our model as the sum of its kinetic and potential energies and explain the two exponents that determine the range of the interactions. Next, in Section \ref{sec3}, we present a detailed  study on the behavior  of the maximal Lyapunov exponents when long--range is applied either to the quadratic (linear LRI) or the quartic (nonlinear LRI) part of the potential. We choose random initial conditions and compare the dynamics and statistics of these cases computing the pdfs of the sums of their momenta for sufficiently long times.

In Section \ref{sec4}, we examine the value of $q$, and other parameters on which the pdf depends, focusing especially on the thermodynamic limit, where the total energy $E$ and $N$ tend to infinity at fixed specific energies $\varepsilon=E/N$. We thus discover, for $0\leq \alpha_2<1$, a \textit{linear} relation between $q$ and $1/ \log N$, which allows us to extrapolate the value of $q=q_{\infty }$ in the limit $N\rightarrow\infty$. Since we thus end up with values $q_{\infty }>1$, we conclude that $q$--Gaussian pdfs behave as if they were \textit{attractors} and hence that $q$--statistics prevails over BG thermostatistics in that limit. Finally in Section \ref{sec5} we present our conclusions.


\section{The FPU $\beta$--model with different ranges of interaction\label{sec2}} 

Let us consider the famous Fermi--Pasta--Ulam $\beta $--model of a 1--dimensional
lattice of $N$ nonlinearly coupled oscillators governed by the Hamiltonian 
\begin{equation}
\label{FPU}
{\cal H}_{FPU}= \frac{1}{2}\sum_{n=1}^{N} p_n^2 + \sum_{n=0}^N V_2(x_{n+1}-x_n) + \sum_{n=0}^N V_4(x_{n+1}-x_n) ~~,
\end{equation}
involving nearest--neighbor interactions, where $V_2$ and $V_4$ represent the quadratic and quartic functions $V_2(u) =a u^2/2$ and $V_4(u)=b  u^4 /4$. The $p_n,x_n$ are the canonical conjugate pairs of momentum and position variables assigned to the $nth$ particle, with $n=1,2,...,N$ and fixed boundary conditions, i.e. $x_0=x_{N+1}=p_0=p_{N+1}=0$.

In this paper we modify the above classical form of the FPU $\beta$--model by introducing the parameters $\alpha_1 $ and $\alpha_2$, which enter in the linear and nonlinear parts of the equations of motion, to determine the particle interactions that decay with distance as $1/r^{\alpha _1}$ and $1/r^{\alpha _2}$ respectively. In particular, the modified Hamiltonian function that describes the generalized FPU $\beta$--system has the form 
\begin{eqnarray}\label{ham}
{\cal H}_{LRI}=\frac{1}{2}\sum_{n=1}^{N} p_n^2 +  \frac{a}{2\widetilde N_1} \sum_{n=0}^{N} \sum_{m=n+1}^{N+1} \frac{(x_n-x_m)^2}{(m-n)^{\alpha_1}} + \frac{b}{4\widetilde N_2} \sum_{n=0}^{N} \sum_{m=n+1}^{N+1} \frac{(x_n-x_m)^4}{(m-n)^{\alpha_2}} ~~,
\end{eqnarray}
where $a$ and $b$ are positive constants.

Note that there are three ways to introduce long--range interactions
in our model: (a) only in the quadratic potential $V_2$, (b) only in the quartic potential $V_4$ and (c) both in $V_2$ and $V_4$. Case (b) was the one studied in \cite{CTB} and gave the results mentioned above, where a ``phase transition'' occurs between $q$--statistics and BG thermostatistics near the value $\alpha_2=1$ that separates the short term $\alpha_2>1$ from the long term $0\leq\alpha_2<1$ interaction range.

As explained above, the critical value $\alpha_1 = \alpha_2= 1$ is expected to determine the crossover between long and short--range interactions. When $\alpha_i <1$, $i=1,2$ the interactions are long range, with the lower bound $\alpha_i =0$ signifying that each particle interacts equally with all others, exactly as in a fully connected network. In contrast, when $\alpha_i > 1$, $i=1,2$, the interactions are short--range and in the limit $\alpha_i \rightarrow \infty $ only the nearest neighbor terms survive in the sums and the classical form of the FPU $\beta $--Hamiltonian is recovered.
  
The rescaling factors $\widetilde N_i$, $i=1,2$ in (\ref{ham}) are given by the expression 
\begin{eqnarray}\label{factor}
{\widetilde N_i}(N,\alpha_i)  \equiv    \frac{1}{N} \sum_{n=0}^{N} \sum_{m=n+1}^{N+1}  \frac{1}{(m-n)^{\alpha_i}}, ~~(i=1,2) 
\end{eqnarray}
and are necessary for making the Hamiltonian extensive. Indeed, without this factor the sums of  $V_2$ and $V_4 $ in (\ref{ham}) would increase as $O(N^2)$ in the thermodynamic limit, thus rendering the kinetic energy (which grows like $N$) irrelevant \cite{CTB}. Notice that $\widetilde N_i \simeq 1$ in the limit $\alpha _i\rightarrow \infty $, and thus for large $N$ Hamiltonian (\ref{ham}) reduces to Hamiltonian (\ref{FPU}).

\section{Conditions for weak chaos and $q$--thermostatistics\label{sec3}}

\subsection{Linear versus nonlinear long--range interactions\label{sus}}

It has been known for some time (see e.g. \cite{Ruffo98,AnteneodoTsallis1998}) that Hamiltonian systems possessing LRI display a more organized behavior in the thermodynamic limit. It is also well 
established that the maximal Lyapunov exponent $\lambda $ of the FPU $\beta$--system
converges to a positive constant in the thermodynamic limit, that depends only on the
coupling constant $b$ and  the system's specific energy $\varepsilon $. 
What happens, however, when linear and/or nonlinear interactions between distant particles are taken into account? Choosing to work on the FPU $\beta$--model in the present paper, provides the advantage of making a direct comparison between linear and nonlinear LRI and allows us to examine in detail their effect on the system's dynamics.

Note that system (\ref{ham}) can be studied in the presence of only linear LRI by taking $\alpha _2 \rightarrow \infty $ and letting $\alpha _1$ act as a free parameter that controls the range of interaction. This means that the potential $V_4$ in that case is equivalent to the one used in the classical FPU $\beta$--model. By contrast, if we wish to study the effect of nonlinear LRI alone, we take $\alpha _1 \rightarrow \infty $ and $\alpha _2$ becomes the free parameter.

\begin{figure}
\begin{center}
\includegraphics[width=0.25\linewidth]{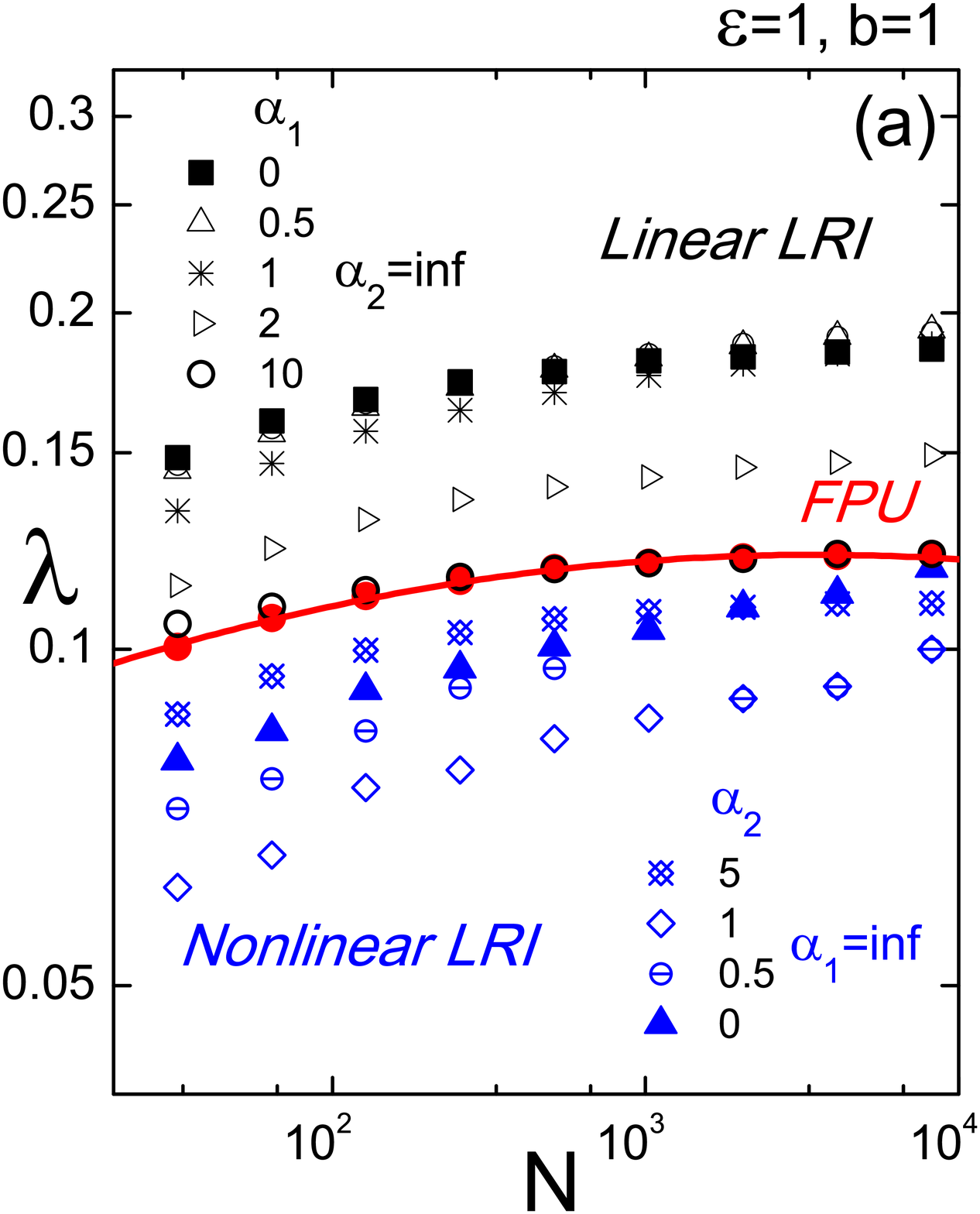}
\includegraphics[width=0.25\linewidth]{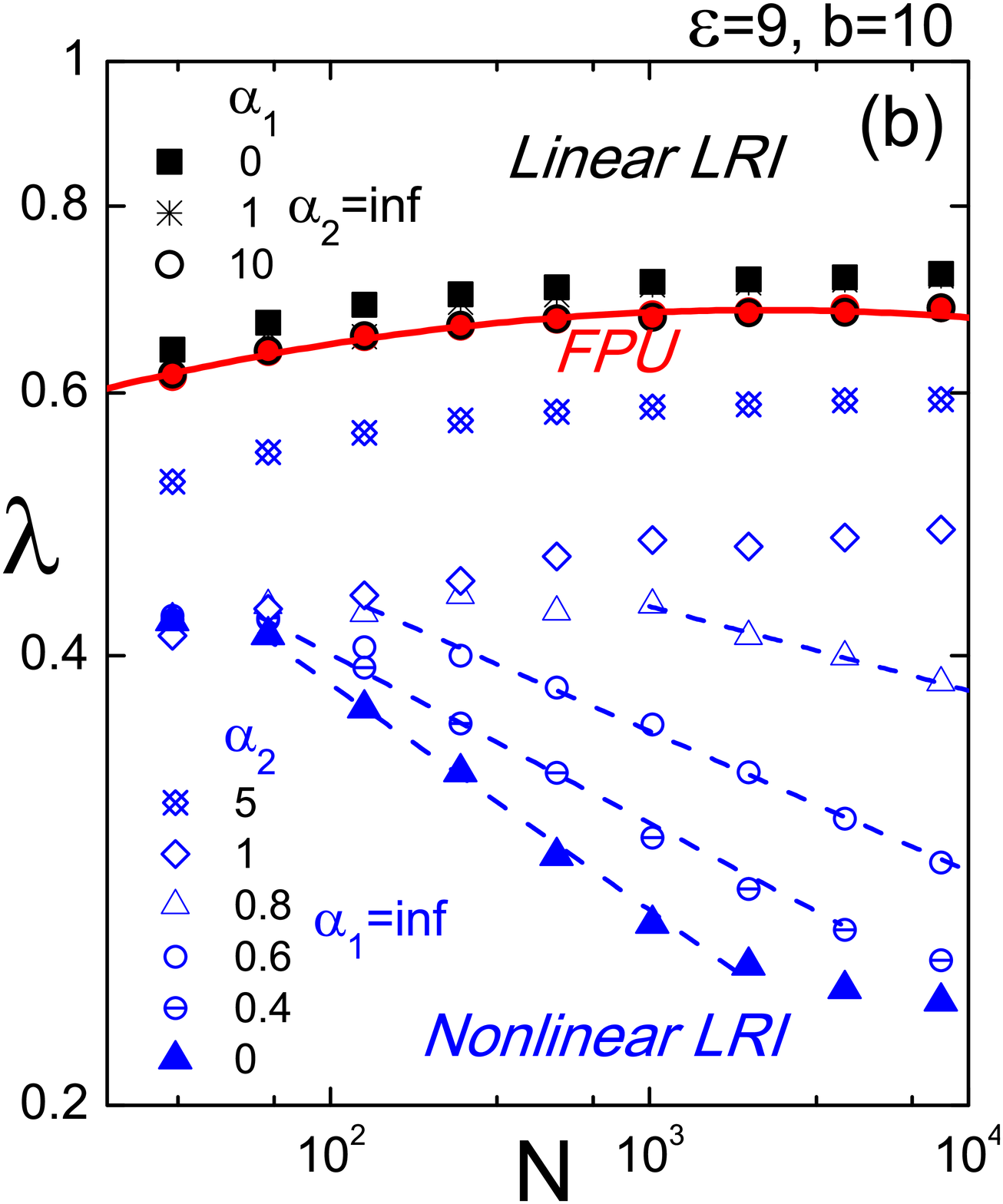}
\includegraphics[width=0.5\linewidth]{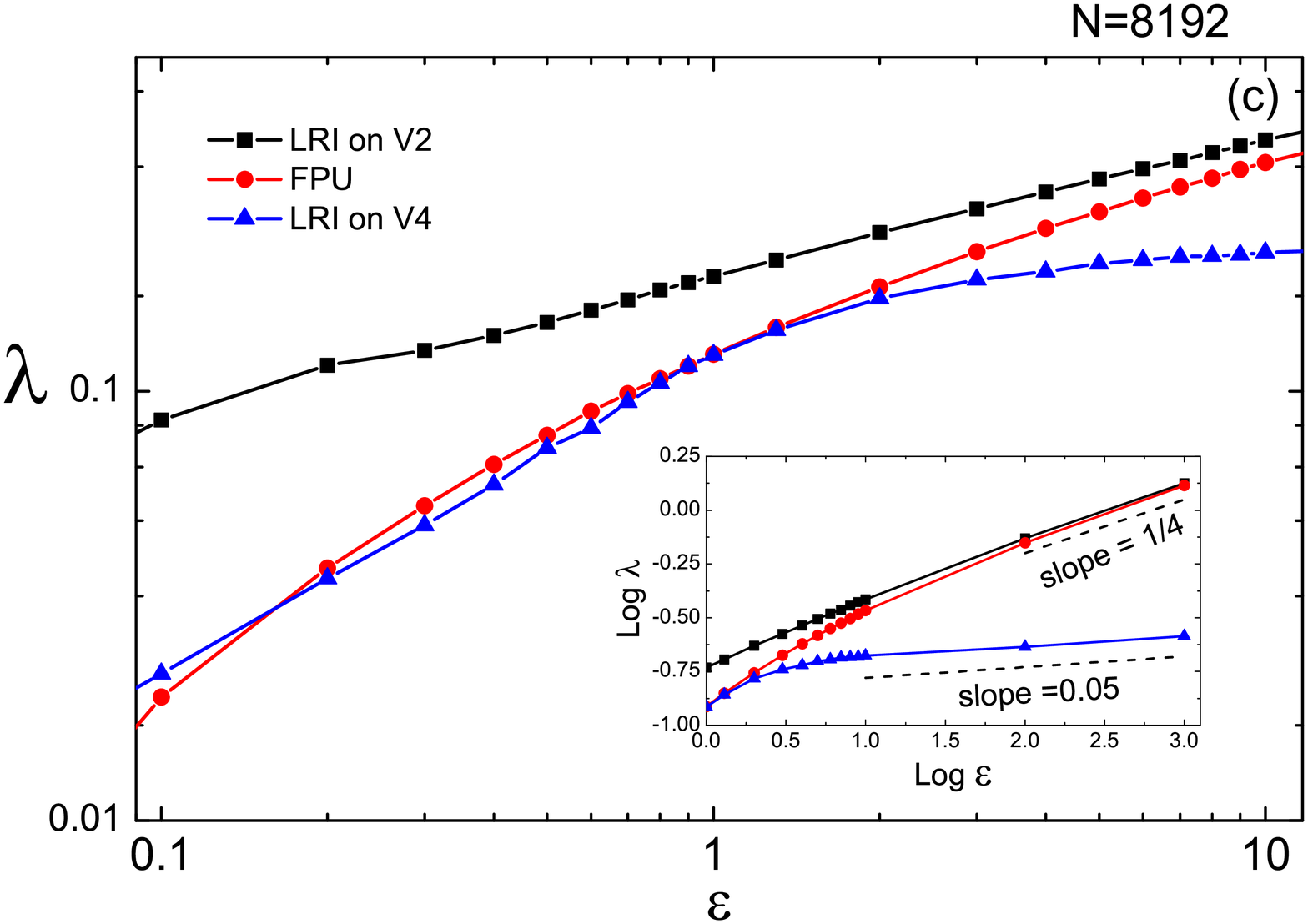}
\end{center}
\caption{Log--log plots of the MLE: (a) For increasing $N$, with $b=1$ and $\varepsilon=1$, (b) as $N$ increases with $b=10$ and $\varepsilon=9$ and (c) as a function of $\varepsilon$ at $N=8192$, $b=1$, for 3 cases: An upper curve of black squares for LRI on $V_2$ only with $(\alpha_1,\alpha_2)=(0,\infty)$, the middle one of red circles for the classical FPU case with $(\alpha_1,\alpha_2)=(\infty,\infty)$ and a lower one of blue triangles for LRI on $V_4$ only for $(\alpha_1,\alpha_2)=(\infty,0)$.
}
\label{LE}
\end{figure}

Let us now display in Fig.\ref{LE} the behavior of the maximal Lyapunov 
exponent $MLE=\lambda $, in the above cases, first in terms of the system size $N$ and then as a function of the specific energy $\varepsilon$ ($a=1$ in every case).  These values tend to stabilize and converge to a constant value for times greater than $10^4$, therefore we stopped our simulations at $t=10^6$.
Panels (a) and (b) are for  $b=1$, $\varepsilon =1$ and $b=10$, $\varepsilon =9$ respectively. Both of them include values $\lambda _{FPU}$ of the classical FPU $\beta$--model as a point of reference, which separate the two LRI cases: Below $\lambda _{FPU}$ we find the maximal Lyapunov exponents $\lambda_{V4}$ of the nonlinear LRI case, while above we encounter the $\lambda_{V2}$ exponents. Note that the longer the range of interaction, the higher the  $ \lambda_{V2} $ values. For $\alpha_1=\alpha_2 =10$ the $\lambda_{V2} $ and $\lambda_{V4} $ curves collapse to the $\lambda _{FPU}$ values.

In Fig.\ref{LE}(a) the MLEs $\lambda_{V2}$ and $\lambda_{V4}$ grow very slowly and  even tend to saturate as $N\rightarrow \infty $. By contrast, in Fig.\ref{LE}(b) for $\varepsilon=9$ and $b=10$, this tendency is reversed in the case of the $\lambda_{V4}$ exponents which are seen to decrease with $N$. The reason this is not observed in Fig.\ref{LE}(a) is because it requires high $b \varepsilon $ values, as pointed out already in \cite{CTB}. An additional remark is that,  for $\alpha_2=0$ the exponents $\lambda_{V4}$ of Fig.\ref{LE}(b) seem not to vanish for $N\rightarrow \infty $, but tend to saturate at a positive value. Only when the quadratic part $V_2$ ($a=0$ in (\ref{ham})) is completely eliminated from the Hamiltonian, the Lyapunov exponents continue to fall to zero as $N$ keeps increasing (see \cite{Bagchi} for a detailed numerical study of this issue). 

So, what are the $b \varepsilon$ values that yield a power--law decrease of $\lambda_{V4}$ vs. $\varepsilon$? When is the system weakly chaotic and why? To find out we have computed the MLEs at various specific energies, keeping the parameter $b=1$ fixed and the number of particles $N=8192$. As Fig.\ref{LE}(c) clearly shows, linear LRI (represented by the upper curve of squares) make the system much more chaotic than the classical nearest neighbor FPU case (represented by the middle curve of circles). We believe that this is due to the fact that the implementation of LRI on the linear part of the Hamiltonian results in a `compression' of the phonon band $\omega_k = 2 \sin \frac{k \pi }{2(N+1)}$ of the nearest neighbor case ($\alpha_1=\infty$) from the interval $[0,2]$ to a single point with frequency $\Omega =\sqrt{2(N+2)/(N+1)}$, as $\alpha_1$ tends to zero. This suggests that no sizable region of quasiperiodic tori exists to sustain regular motion, while the periodic oscillations of the lattice (with frequency $\Omega$) that become unstable due to the presence of nonlinear terms should have large scale chaotic regions about them that dominate the dynamics in phase space.
 
On the other hand, when LRI apply only to the nonlinear part of the Hamiltonian and the harmonic terms are of the nearest neighbor type it is interesting to compare the chaotic behavior of the system with that of the classical FPU model. As Fig.\ref{LE}(c) clearly demonstrates, the corresponding MLE curves are very close to each other when $\varepsilon$ is small, but begin to deviate considerably for $\varepsilon >1$. Indeed, the application of nonlinear LRI is characterized by much weaker chaos in the limit $\varepsilon\rightarrow \infty $ as its MLE behaves like $\lambda_{V4} \sim \varepsilon^{0.05}$, in contrast with the FPU model whose corresponding MLE grows a lot faster, as $\lambda_{FPU} \sim\varepsilon ^{1/4}$ (see \cite{casetti} for an analytical derivation). It is also interesting to note that in this limit the exponents $\lambda_{V2}$ and $\lambda_{FPU}$ become indistinguishable.

Remarkably, the above picture of the maximal Lyapunov exponent $\lambda_{V4} $ slowing down its increase as the specific energy grows (see the triangles in Fig.\ref{LE}(c)) is accompanied by the emergence of $q$--Gaussian distributions in the momenta associated with the presence of weakly chaotic behavior. In the next subsection we examine this phenomenon more carefully as we concentrate our study on the statistical aspects of the LRI models.

\subsection {Emergence of q--Gaussian distributions}
Besides this striking difference of the level of chaoticity in the above two (short and long--range) situations, there is also a remarkable difference in their statistics, as we now explain.

 The pdfs we study correspond to the momenta $p_1(t),\ldots,p_N(t)$ of orbits starting from a uniform distribution at $t=0$. In particular, these are time averaged pdfs which are evaluated at discrete times $t_j>t_0,j=1,2,\dots$, where $t_0$ is the time at which the kinetic energy has stabilized. The step $\tau =t_{j+1}-t_j$  between these discrete times should be appropriately chosen so as to avoid possible correlations. 
We then assign to the $i$--th momentum band (i.e. $i$--th column of our histograms)
the number of times where the momentum for each single particle falls 
in.

\begin{figure}
\begin{center}
\includegraphics[width=0.58\linewidth]{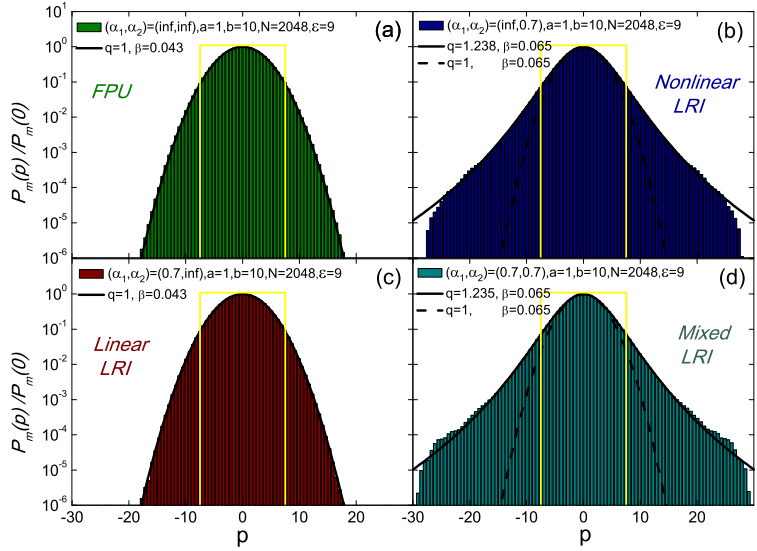}
\end{center}
\caption{ The momentum distributions for $N=2048$ particles for the system (\ref{ham}). 
The upper panels show the cases: $\alpha_1 \rightarrow \infty$, $\alpha_2 \rightarrow \infty$, i.e. FPU (left) and $\alpha_1 \rightarrow \infty$, $\alpha_2=0.7$ (right). Lower panels
show: $\alpha_1=0.7$, $\alpha_2 \rightarrow \infty$ (right) and $\alpha_1=\alpha_2=0.7$ (right). The yellow lines correspond to the uniform distribution, from which the momenta where randomly extracted.}
\label{first}
\end{figure}

\begin{figure}
\begin{center}
\includegraphics[width=0.58\linewidth]{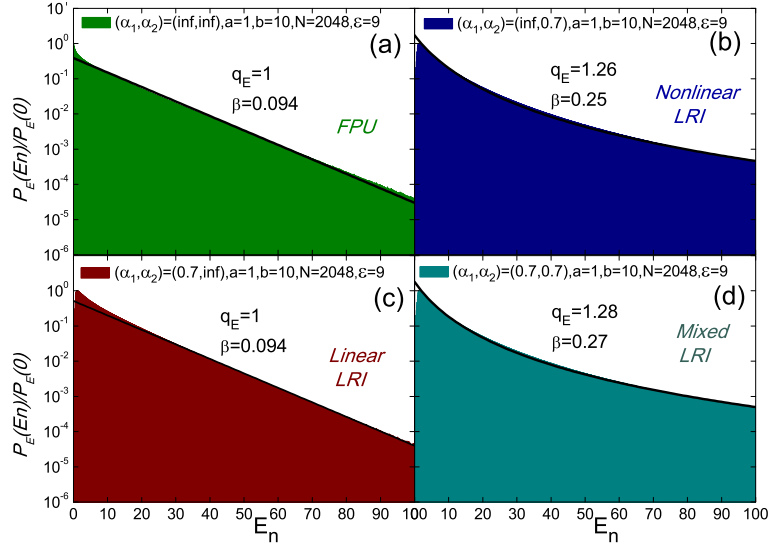}
\end{center}
\caption{  The particle energy distributions for the 4 cases of Fig. \ref{first}; { notice that the fittings (black continuous curves, along the upper bounds of the numerical histograms) exhibit BG distributions ($q_E=1$) when the quartic interactions are short--ranged, and $q_E$-exponential distributions ($q_E>1$) when they are long--ranged}. \label{klk}}
\end{figure}

 In the four panels of Fig.~\ref{first} typical momentum histograms are shown, which correspond to the four representative cases we study, as different combinations of short and long--range interactions are applied on $V_2$ and $V_4$.  More specifically, in panels (a) and (c) a classical Gaussian shape is observed, either under purely short--range interactions or when LRI apply only to the quadratic part, by setting $\alpha_1=0.7$ and $\alpha_2\rightarrow \infty $ in the Hamiltonian (\ref{ham}).  Instead in the panels (b) and (d) a clear $q$--Gaussian shape emerges when long--range applies to the quartic interactions, independently of the type of interactions in the quadratic part, i.e. for $\alpha_1\rightarrow \infty $, $\alpha_2=0.7$ and $\alpha_1=0.7$, $\alpha_2=0.7$ in (\ref{ham}) respectively.  These plots have been evaluated at $\tau =2$ time steps within the interval $[10^5, 5\cdot 10^5]$.
  Furthermore, as we see from the plots of Fig. \ref{klk}, the distributions of the time averaged individual particle energies (i.e. $E_n= {1\over 2} p_n^2 + \sum_m V(x_n - x_m)$, which are evaluated after the same time interval as in Fig.~\ref{first} and for the same initial conditions,  show very similar behavior to the momenta distributions: $q$--Exponentials appear when the LRI are nonlinear.

In practice, we have employed an algorithm which uses the least squares method to calculate $q$, determines the intercept and estimates $\beta $ from the slope of the resulting straight line. Dividing then the $q$ interval [1,3] into 1000 possible values, we apply the least squares method to all of them. The appropriate $q$ value is chosen as the one corresponding to the minimum standard error and is estimated with an accuracy of at least 3 digits. On the level of $q$-statistics the $\beta$ parameter entering our distributions corresponds to an inverse effective temperature characterizing the width of the distribution. In other words, if this fact is viewed within a thermostatistical framework where an entropy and a partition function can be defined, one expects $\beta=1/kT$, where $T$ corresponds to the temperature and $k$ is Boltzmann constant \cite{Tsallis2009}.

As is evident from these results, the mechanism of LRI drives the system's behavior away from BG statistics, only if the quartic potential is long--range. Instead, when LRI apply only to the quadratic part, purely Gaussian pdfs are obtained. 

In what follows, we examine which of the system's fundamental parameters affect the shape of the $q$--Gaussian pdfs and how the interaction range of the quadratic part of the potential influences the system's behavior.

\section{Variation of $q$ for different ranges and system parameters\label{sec4}}

As is well--known, $q$--Gaussian distributions are often associated with weak chaos and are linked to QSS which persist for very long times, until the system achieves energy equipartition at complete thermalization.  In such cases, there always exists a value of time $t_c$ beyond which the system passes to a strongly chaotic state characterized by $q=1$ and BG thermostatistics\cite{ABB2011,bountis_book}. This is also what happens with all weakly chaotic states in the LRI FPU-$\beta$ model and that is why we refer to them as QSS. Thus, our system is different in this regard from the $\alpha$--XY model studied in \cite{CirtoAssisTsallis2013}, for which $q$-Gaussian distributions are found to persist even in the so-called BG regime.

\begin{figure}
\begin{center}
\includegraphics[width=0.45\linewidth]{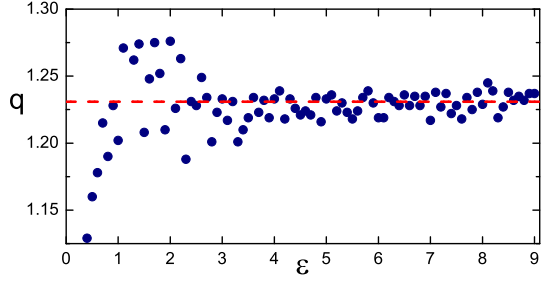}
\end{center}
\caption{The dependence of $q$ on $\varepsilon$, when LRI are applied to $V_4$ with $\alpha_2=0.7$, $\alpha_1=\infty$, $a=1$, $b=10$ and $N=2048$.  The dashed line corresponds to a mean value of about $q=1.23$, which is close to the $q_E=1.26$ of Fig.\ref{klk}(b), also evaluated at $\varepsilon=9$. It is not transparent, however, if this is coincidental or due to some deeper explanation related to the fact that the particle energy includes the particle kinetic term. \label{qvse}}
\end{figure}

\begin{figure}
\begin{center}
  \includegraphics[width=0.55\linewidth]{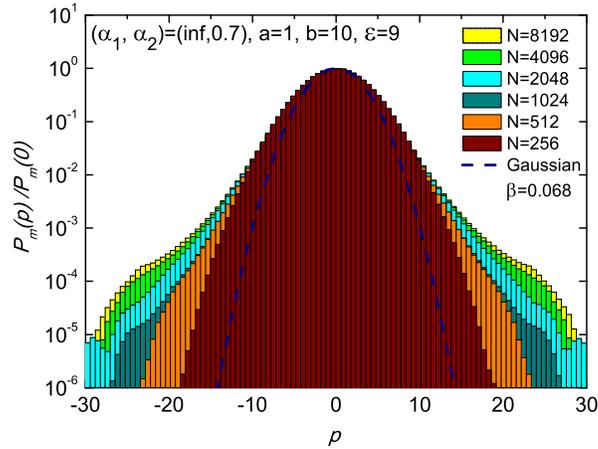}
 \end{center} 
\caption{Momentum distributions for the system with $b=10, \varepsilon=9, \alpha _2=0.7$ and various $N$ values. Note how the pdfs are described by a $q$--Gaussian of higher index $q$ as $N$ grows. More specifically, $q$ ranges from $1.17$ for $N=512$ until $1.25$ for $N=8192$.}
\label{Ndep}
\end{figure}

\begin{figure}
\begin{center}
\includegraphics[width=0.38\linewidth]{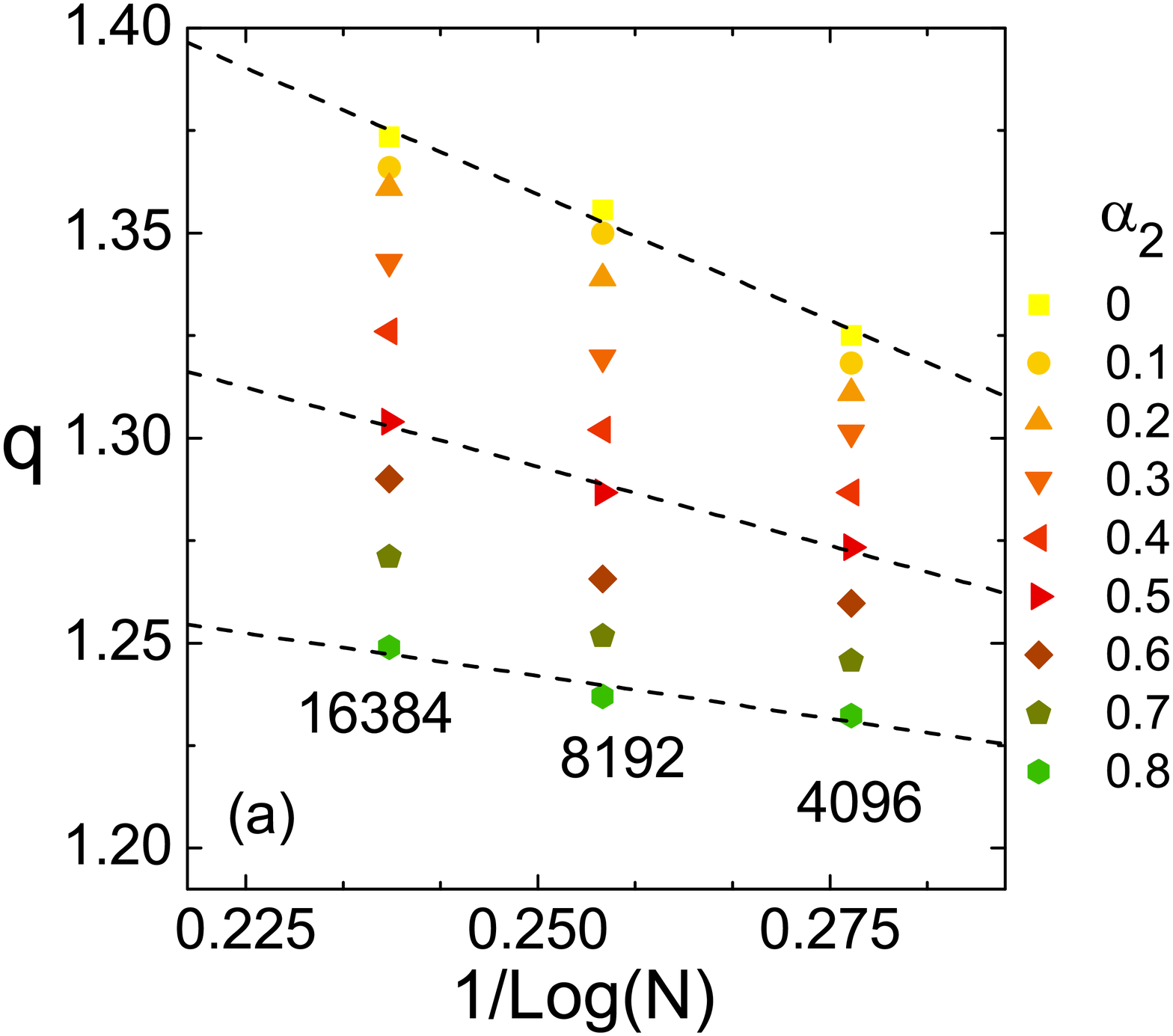}
\includegraphics[width=0.38\linewidth]{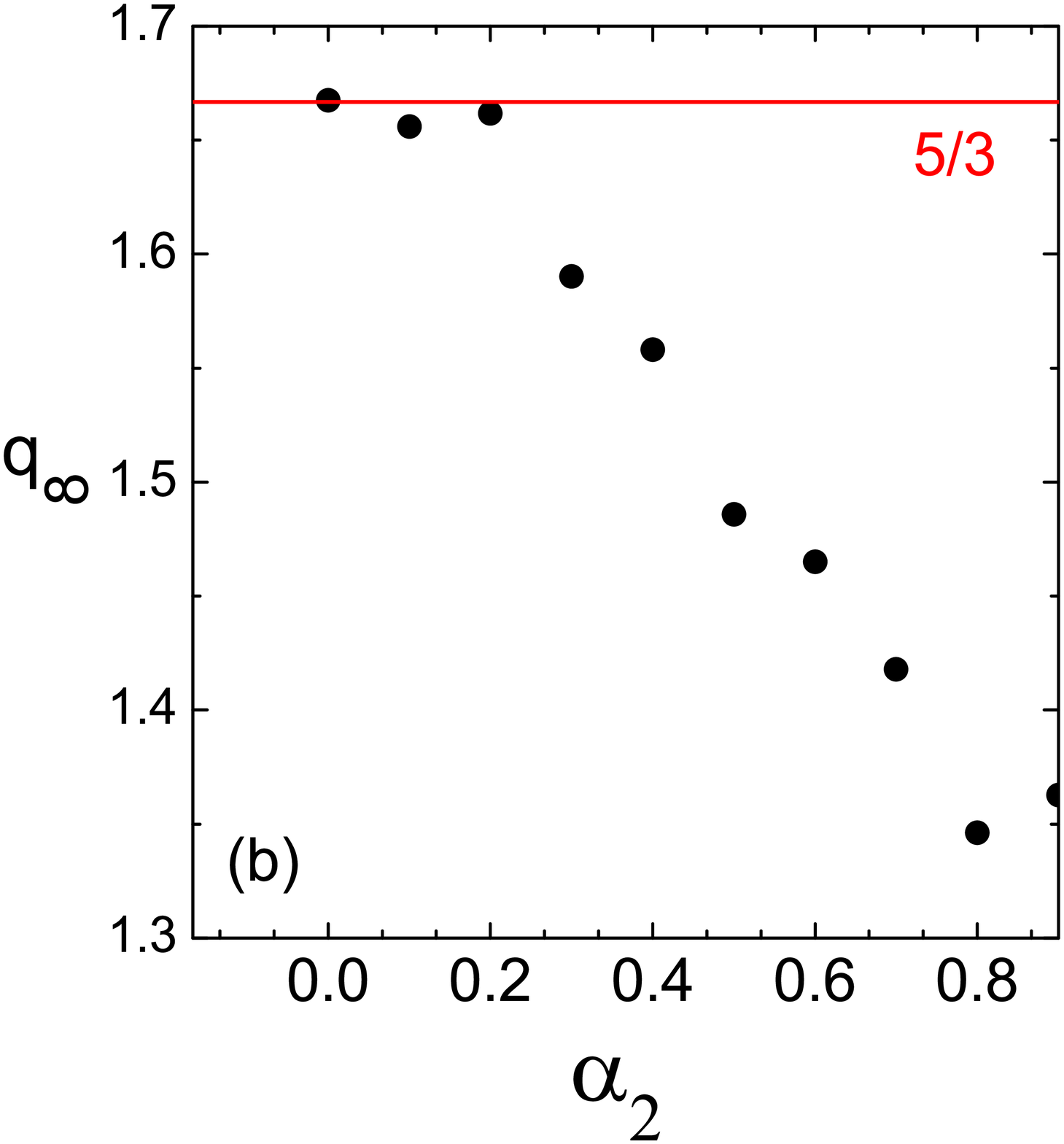}
\end{center}
\caption{ (a) The linear dependence of $q$ on $1/log N$ for $N=4096, 8192, 16384$ depicted here provides an estimate for $q_{\infty}$ in the thermodynamic limit, { as $\alpha_2$ changes, with $\alpha_1=\infty$}. (b) The values of $q_{\infty }$ are plotted here versus $\alpha _2$. We have not included $\alpha_2>0.8$ in the above results due to the `noisy' behavior of $q$ in the neighborhood $\alpha_2=1$. Nevertheless, for $\alpha_2$ above 1.4, we definitively obtain $q=1$. ($\varepsilon=9$ in both panels.)  \label{uni}}
\end{figure}

Our main purpose here is to investigate numerically the dependence of $q$ on the system size
$N$, the specific energy $\varepsilon$ and the coupling constant $b$ of the 
Hamiltonian (\ref{ham}). Let us mention at the outset that the parameters $\varepsilon$ and $b$ are not independent. Indeed, a simple rescaling of the Hamiltonian shows that the relevant parameter is $b \varepsilon$.

Let us plot in Fig.\ref{qvse} the dependence of $q$ on $\varepsilon$ for $ N=2048$ and $\alpha_1\rightarrow\infty$. 
It turns out that $q$ fluctuates around 1.23 and displays a greater tendency to converge as the specific energy increases.
It is important to note that when linear LRI are  added to the nonlinear LRI the motion is still weakly chaotic and the index $q$ of the associated pdfs remains unaffected.

On the other hand, when the system size increases, the value of $q$ no longer remains a constant but also increases with $N$. From Fig.~\ref{Ndep} it becomes evident that the corresponding $q$--Gaussian representing the statistics of the model spreads as $N$ grows. Thus, choosing $\alpha_2 =0.7$, $a=1, b=10$ and $\varepsilon =9$ we find that the momentum histogram for low values of $N$ is very close to a Gaussian, as shown in Fig.~\ref{Ndep} for $N=256$. It then deviates for $N=512$ to a $q$--Gaussian with $q(N=512)=1.17$, which further increases to $q(N=1024)=1.19$ and so on, as the weakly chaotic properties of the dynamics become more evident.  

\subsection{Asymptotic behavior of $q$ in the limit $N\rightarrow \infty$}

Extrapolating the value of $q$ in the limit $N\rightarrow \infty $, we can now estimate the asymptotic value  $q=q_{\infty }$ and also vary $\alpha_2$ to determine the dependence of $q_{\infty }$ on the interaction range applied to the quartic part of the potential at the thermodynamic limit. To this end, we consider a given value of $\alpha_2<1$ and systematically calculate the $q$ dependence on $N$. In  Fig.~\ref{uni}(a) we plot these $q$ values versus $1/ \log N$ and find that their dependence is accurately described by the following expression: 
\begin{eqnarray}\label{qvsN}
q(N,\alpha_2) = q_{\infty } (\alpha_2) - c(\alpha_2)/ \log N  ~~ ,
\end{eqnarray}
where $c(\alpha_2 )$ is some constant. Each of the data in  Fig.~\ref{uni}(a) has been plotted after performing 3 independent realizations of the momentum distributions and taking their average in the time window $[10^5, 5\cdot 10^5]$. 

This appears to be an important result because it shows that the $q_{\infty } (\alpha_2)$ obtained from Fig.~\ref{uni}(a) by the intercept of the straight line Eq.~(\ref{qvsN}) with the vertical axis (as $N\rightarrow\infty$) is larger than 1, which implies that the $q$--Gaussians are attractors in that limit. Next, plotting $q_{\infty } (\alpha_2)$ vs. $\alpha_2$ in Fig.~\ref{uni}(b), we observe that it starts from $5/3$ for $\alpha _2 =0$, and then, after about $\alpha_2 =0.2$, falls linearly towards 1. In particular, for $0.2 \leq \alpha_2 \leq 0.8$ the values of $q_{\infty } (\alpha_2)$ decrease as $q_{\infty } (\alpha_2) = 1.79 -0.475\alpha_2$.  Concerning the left limit, we need to recall that $q$-Gaussians have finite variance for $q$ up to $5/3$. Therefore, this value constitutes a natural candidate for quartic infinitely long interactions, i.e. $\alpha_2=0$, as evidenced also by numerical results obtained in \cite{CTB,Tsallis2009}. 

Note that the value of $q$ reaches unity at $\alpha_2 =1.5$ and not at the expected $\alpha_2 =1$ threshold between short and long--range interactions. This is a very interesting phenomenon and may be explained by the fact that $q$ takes a very long time to converge to 1 over the range $1 \leq \alpha_2 \leq  1.4$.

\section{Conclusions\label{sec5}}

In the present paper a generalization of the 1-dimensional Fermi-Pasta-Ulam $\beta$--model was studied, where two non--negative exponents  $\alpha_1$ and $\alpha_2$ are introduced in the quadratic $V_2$ and quartic $V_4$ part of the potential to control the range of interactions. The role of long--range interactions on the system's dynamical properties as well as its statistical behavior were examined in detail. In particular we concluded that only when LRI apply on $V_4$ and at high enough $b \varepsilon$ values: (a) the maximal Lyapunov exponent  $\lambda_{V4}$ decreases as a power-law with $N$  \cite{CTB} 
and increases very slowly, as  $\lambda_{V4} \sim  \varepsilon^{0.05}$ ($b=1$) with the specific energy, while at the same time (b) $q$--Gaussian pdfs of the momenta appear with $q>1$. Both of these results indicate that LRI on $V_4$ is a necessary condition for what we call weak chaos in the FPU $\beta$--model, especially at high energies.

On the contrary, when LRI are applied to the harmonic part of the potential a much stronger type of chaos is encountered if the nonlinear interactions are short--range. This is especially evident at low energies, indicating that the transition to large scale chaos occurs at much lower levels than in the classical FPU case. The corresponding MLE=$\lambda_{V2}$ tends to saturate as a function of $N$, while, for small $\varepsilon$, $\lambda_{V2}$ is much higher than the MLE=$\lambda_{FPU}$ of the classical FPU model, with $\lambda_{FPU}$ tending to $\lambda_{V2}$ from below as $\varepsilon\rightarrow\infty$. All this is related to momentum pdfs of the purely Gaussian type ($q=1$) and is associated with strong chaos and BG thermostatistics.

We also focused on the value of $q$ in the momentum pdfs, when the main parameters of the problem vary. It turns out that $q$ changes with $N$ and $\alpha _2$ and not with $\alpha _1$. On the other hand, when $0\leq \alpha_2<1$ we find a \textit{linear} relation between $q$ and $1/ \log N$, which allows us to extrapolate the value of $q$ to $q_{\infty}>1$ at $N\rightarrow\infty$. This is important because it suggests that under these conditions BG thermostatistics no longer holds and $q$--Gaussian and $q$--exponential pdfs with $q>1$ describe the true statistics in the thermodynamic limit. 


\paragraph{Acknowledgments}
We are especially indebted to the referees for their many detailed comments and remarks that helped us improve significantly our manuscript. One of us (C.T.) gratefully acknowledges partial financial support by the Brazilian Agencies CNPq and Faperj, and by the John Templeton Foundation (USA). All of us acknowledge that this research has been co-financed  by the European Union (European Social Fund--ESF) and Greek national funds through the Operational Program `Education and Lifelong Learning' of the National Strategic Reference Framework (NSRF) - Research Funding Program: {\it Thales. Investing in knowledge society through the European Social Fund}.


\begin{thebibliography}{99}


\bibitem{Ruffo95} Antoni M and Ruffo S, 1995 {\it Phys. Rev. E} \textbf{52}  2361-2374. 

\bibitem{Ruffo98} Latora V, Rapisarda A and Ruffo S, 1998 {\it  Phys. Rev. Lett}  \textbf{80} 692.

\bibitem{AnteneodoTsallis1998} Anteneodo C and Tsallis C, {\it Phys. Rev. Lett.} {\bf 80} 5313.



\bibitem{Dauxois2002} Dauxois T, Latora V, Rapisarda A, Ruffo S and Torcini A, 2002
Lecture Notes in Physics, edited by  Dauxois T, Ruffo S, Arimondo E, Wilkens M,  Vol. 602 p. 458.

 \bibitem{Tarasov06} Tarasov V E and Zaslavsky G M 2006 {\it Commun. Nonlinear Sci. Numer. Simul.}  \textbf{11} 885-898. 

\bibitem{Ginelli1}  Takeuchi K A, Chat{\'e} H, Ginelli F, Politi A and Torcini A
2011 {\it Phys. Rev. Lett.} \textbf{107} 124101.

\bibitem{Ginelli2} Ginelli F, Takeuchi K A, Chat{\'e} H, Politi A and Torcini A  
2011 {\it Phys. Rev. E} \textbf{84} 066211.

\bibitem{Thanos}  Manos Th and Ruffo S 2011 {\it Trans. Theor. and Stat. Phys. } \textbf{40} 360-381.

\bibitem{Bagchi}   Bagchi D and  Tsallis C { Sensitivity to initial conditions of $d$-dimensional long--range interacting Fermi-Pasta-Ulam model: Universal scaling}, 1509.04697 [cond-mat.stat-mech].


\bibitem{CTB}  Christodoulidi H, Tsallis C and  Bountis T 2014 \textit{EPL} {\bf 108} 40006.

\bibitem{GALI} Skokos H, Bountis T and Antonopoulos Ch 2007 \textit{Physica D} {\bf 231} 30-54.

\bibitem{bountis_book} Bountis T and Skokos H 2012 {\it Complex Hamiltonian Dynamics and Statistics},  Springer Series in Synergetics, Berlin. 









\bibitem{Leo1} Leo M,  Leo R A and Tempesta P 2010 {\it J. Stat. Mech.} P04021. 


\bibitem{Leo2} Leo M,  Leo R A, Tempesta P and Tsallis C 2012 {\it Phys. Rev. E} {\bf 85} 031149.


\bibitem{Leo3} Leo M,  Leo R A and Tempesta P 2013 {\it Annals Phys.} {\bf 333} 12-18.


\bibitem{PluchinoRapisardaTsallis2007} Pluchino A, Rapisarda A and Tsallis C
2007 {\it Europhys. Lett.} {\bf 80} 26002.


\bibitem{PluchinoRapisardaTsallis2007b} Pluchino A, Rapisarda A and Tsallis C 2008 {\it Europhys. Lett.} {\bf 83} 30011.



\bibitem{CirtoAssisTsallis2013} Cirto L J L, Assis V and Tsallis C {\it Physica A} {\bf 393} (2014)   286-296.


\bibitem{uma1} Umarov S, Tsallis C and Steinberg S 2008 {\it Milan J. Math.} {\bf 76}  307-328.


\bibitem{uma2}  Umarov S, Tsallis C, Gell-Mann M and Steinberg S 2010 {\it J. Math. Phys.} {\bf 51} 033502.


\bibitem{uma3} Hahn M G, Jiang X X and Umarov S {\it J. Phys. A} {\bf 43}, (16) (2010) 165208.
\bibitem{uma4} Hilhorst H J 2010 {\it J. Stat. Mech.} P10023. 
\bibitem{uma5} Jauregui M, Tsallis C and Curado E M F 2011 {\it J. Stat. Mech.} P10016.
\bibitem{uma6} Jauregui M and Tsallis C, 2011 {\it Phys. Lett. A} {\bf 375} 2085-2088.
\bibitem{uma7} Plastino A and Rocca M C, 2012 {\it Milan J. Math.} {\bf 80} 243-249.
 
\bibitem{TsallisPlastinoAlvarezEstrada2009}  Tsallis C,  Plastino A and Alvarez-Estrada R F , 
2009 {\it J.~Math. Phys.} {\bf 50} 043303.


\bibitem{Gibbs1902}  Gibbs J W, {\it Elementary Principles in Statistical Mechanics -- Developed with Especial Reference to the Rational Foundation of Thermodynamics} (C. Scribner's Sons, New York, 1902; Yale University Press, New Haven, 1948; OX Bow Press, Woodbridge, Connecticut, 1981).


\bibitem{ABB2011} Antonopoulos Ch, Bountis T and Basios V 2011 {\it Physica A} {\bf 390} 3290-3307.

\bibitem{Tsallis2009}  Tsallis C, {\it Introduction to Nonextensive Statistical Mechanics - Approaching a Complex World} (Springer, New York, 2009).

\bibitem{Tsallis1988}  Tsallis C, 1988 {\it Stat. Phys.} {\bf 52} 479  [First appeared as preprint in 1987: CBPF-NF-062/87, ISSN 0029-3865, Centro Brasileiro de Pesquisas Fisicas, Rio de Janeiro].

\bibitem{GellMannTsallis2004}   Gell-Mann M and  Tsallis C, eds., {\it Nonextensive Entropy - Interdisciplinary Applications} (Oxford University Press, New York, 2004).

\bibitem{Tsallis2014} Tsallis C 2014 Contemporary Physics {\bf 55}  179-197. 

\bibitem{FPU}  Fermi E,  Pasta J and Ulam S 1955 Los Alamos, Report No. LA-1940. See also: Newell A C, 
{\it Nonlinear Wave Motion},  Lectures in Applied Mathematics {\bf 15} 143-155 (Amer.~Math.~Soc., Providence, 1974); Berman G P and Izrailev F M, 2005 {\it Chaos} \textbf{15} 015104.

\bibitem{Ruffo15}  Miloshevich G,  Nguenang J P, Dauxois T, Khomeriki R and Ruffo S 2015
{\it Phys. Rev. E}  {\bf 91} 032927.

\bibitem{casetti}  Casetti L,  Livi R and Pettini M 1995 {\it Phys. Rev. Lett.} \textbf{74} 375--378.







\end{thebibliography}
\end{document}